# Post-injection normal closure of fractures as a mechanism for induced seismicity

E. Ucar[1], I. Berre[1,2], and E. Keilegavlen[1]

[1] Department of Mathematics, University of Bergen, Bergen, Norway.

[2] Christian Michelsen Research, Bergen, Norway.

Corresponding author: Eren Ucar (eren.ucar@uib.no)

**Key Points:**

- Normal closure of stimulated fractures after the termination of injection enhances post-injection seismicity.
- Processes are strongly affected by the complex structure of three-dimensional fracture networks.
- Consistent with microseismic data analyses, our simulations show that seismic events occur at the rim of the stimulation region.

**Abstract**

Understanding the controlling mechanisms underlying injection-induced seismicity is important for optimizing reservoir productivity and addressing seismicity-related concerns related to hydraulic stimulation in Enhanced Geothermal Systems. Hydraulic stimulation enhances permeability through elevated pressures, which cause normal deformations, and the shear slip of pre-existing fractures. Previous experiments indicate that fracture deformation in the normal direction reverses as the pressure decreases, e.g., at the end of stimulation. We hypothesize that this normal closure of fractures enhances pressure propagation away from the injection region and significantly increases the potential for post-injection seismicity. To test this hypothesis, hydraulic stimulation is modeled by numerically coupling fracture deformation, pressure diffusion and stress alterations for a synthetic geothermal reservoir in which the flow and mechanics are strongly affected by a complex three-dimensional fracture network. The role of the normal closure of fractures is verified by comparing simulations conducted with and without the normal closure effect.

**1 Introduction**

Enhanced Geothermal System (EGS) technology is considered key to unlocking geothermal energy resources because it can allow for the production of geothermal energy resources that are less dependent on initial hydrogeological conditions. EGSs can be created by hydraulically stimulating the reservoir to enhance permeability and achieve commercial flow rates. The elevated pressures activate pre-existing fractures by decreasing the friction resistance, which may lead to shear failure of the fractures depending on the fracture orientation relative to the direction and strength of the background stress anisotropy. Following shearing, the dilation of the fractures occurs in the normal direction of the fracture surfaces, which results in a permanent increase in the fracture permeability. Such a treatment is called shear dilation stimulation (which is also known as shear stimulation, hydroshearing or low-pressure stimulation). As long as the elevated pressures are below the minimum principal stress, shear dilation remains a dominant mechanism for fracture opening [*Pine and Batchelor*, 1984].

Hydraulic stimulation is essential for geothermal development in low-permeability formations, such as crystalline basement rocks. However, excessive induced seismicity is a by-product observed in many EGS projects [*Majer et al.*, 2007]. Concerns related to induced seismicity can lead to the termination of costly projects, which was observed for the Basel geothermal project [*Häring et al.*, 2008], or cause unfavorable public perception [*Majer et al.*, 2007]. These concerns provide additional motivation to identify and understand the controlling mechanisms underlying the stimulation process and incorporate these mechanisms in modeling approaches.

Fluid injection causes elevated pressures inside the fractures and reduces contact forces, resulting in decreased frictional resistance between the fracture surfaces. These changes facilitate shear slip in fractures that are favorably oriented in the anisotropic background stress field. This mechanism is known as effective stress reduction and has been identified as the major cause of induced seismicity in numerous modeling studies [*Bruel*, 2007; *Kohl and Mégel*, 2007; *Rothert and Shapiro*, 2003]. The importance of stress redistribution following the shearing of rock for injection-induced seismicity has been emphasized [*Catalli et al.*, 2013; *Catalli et al.*, 2016], and the stress alteration has been coupled with the fluid flow in several models [*Baisch et al.*, 2010;

*McClure and Horne*, 2011]. As mentioned by *Baisch et al.* [2010], the stress change in the reservoir caused by slip must be included during numerical modeling because the shear stress of the slipped area decreases by the shear slip and the shear stress at the neighboring region is correspondingly increased; hence, slip avalanches can be initiated.

Particular geothermal fields, such as Soultz-sous-Forêts, Basel and Paralana, provide interesting examples of induced seismicity because large-magnitude events occurred in the shut-in period of the hydraulic stimulations, which is the period when fluid injection had been terminated [*Albaric et al.*, 2014; *Evans et al.*, 2005; *Häring et al.*, 2008]. In related modeling studies, post-injection events have been associated with the continuation of pressure diffusion and stress redistribution in the domain [*Baisch et al.*, 2010; *McClure and Horne*, 2011]. In a recent study, *McClure* [2015] examined a scenario with a simplified fracture network that included a series of parallel fractures pressurized differentially at the end of the injection. During the post-injection phase, the fluid backflows through the well from the highly pressurized fractures to the fractures with lower-pressure conditions resulted in an increase in the post-injection seismicity. Although these previous studies provided valuable insights into the physics of the post-injection seismicity, additional effects remain to be investigated.

The seismic recordings of the Basel and Paralana geothermal projects show that the locations of seismic events after the termination of injection are mainly at the outer rim of the previous seismic activity, i.e., the stimulation region [*Albaric et al.*, 2014; *Häring et al.*, 2008; *Mukuhira et al.*, 2017]. Here, we analyze a novel hypothesis that pertains to the mechanism underlying large post-injection events occurring at the boundaries of the stimulation region. The hypothesis is motivated by experiments conducted by *Bandis et al.* [1983] and *Barton et al.* [1985], who demonstrated fracture deformation under different stress conditions. The fracture deformation takes various forms, such as normal closure, opening, shearing and dilation, depending on the stress state of the fractures. The fluid pressure in the fractures has been shown to determine the normal loading on the fractures and thus the void space, or the aperture. Although the elevated pressure increases the aperture during injection, the opposite effect occurs after the injection is terminated. As the fracture starts to close, the aperture decreases, which primarily occurs in the near-well region. Our hypothesis is that a reduction in aperture will act as a post-injection pressure support to advance the front of the elevated fluid pressure and cause seismic events and a corresponding increase in apertures beyond the previously stimulated region, which was observed in the seismic recordings of Basel and Paralana. To qualify the effect of the normal closure of fractures in the post-injection phase, we examine two scenarios with and without the normal closure mechanism included after the termination of the injection. Moreover, we show how the fracture closure mechanism is manifested in a reservoir that is structurally dominated by a complex three-dimensional fracture network.

## 2. Modeling and Simulation Approach

### 2.1 Modeling of Fracture Deformation

The fractured reservoir rock is modeled as a combination of high-permeability fractures and a porous medium surrounding the fractures, i.e., a low-permeability rock matrix. In computational modeling, the treatment of the fractures is crucial because they structurally dominate the physical processes. The fluid flow is dominated by the fractures because their permeability is significantly higher than that of the rock matrix. Moreover, although the rock matrix can be considered an elastic material that linearly deforms [*Jaeger et al.*, 2009], *Bandis et al.* [1983] and *Barton et al.* [1985] showed that the fractures are more deformable than the rock matrix and should not be modeled as a linearly elastic medium. The fracture deformation is instead modeled by the Barton-Bandis closure model, which defines the relationship between the normal loading, $\sigma_n$, and fracture closure, $\Delta E_{rev}$, as follows:

$$\sigma_n = \frac{K'_n \Delta E_{rev}}{1 - \frac{\Delta E_{rev}}{\Delta E_{max}}} \qquad (1)$$

where $K'_n$ is the initial normal stiffness per area and $\Delta E_{max}$ is the maximum possible closure. Following *Hicks et al.* [1996], the normal stiffness of the fractures is considered to be constant and equal to the initial normal stiffness. Considering that the fracture network is under compression and stress is defined as positive for compression, the normal loading in a pressurized fracture is defined as the 'effective normal loading', which is equal to the difference between the normal stress on the fracture and the pressure inside. Thus, the hyperbolic relationship between the effective normal stress and the fracture closure (equation (1)) is used to determine the reversible deformation depending on the pressure inside the fracture.

The pressurization of the fractures reduces the frictional resistance and ultimately leads to fracture slip in the shear direction. In the present model, the Mohr–Coulomb criterion is used to estimate the shear failure of the fractures according to a friction coefficient initially corresponding to the static friction, $\mu_s$. After a fracture element has failed, its corresponding friction coefficient is reduced to a dynamic friction coefficient, $\mu_d$. The shear stress is released by a permanent shear displacement along the fracture surfaces that are exposed to shear failure. The shear displacement can be calculated by the excess shear stress concept [*Rahman et al.*, 2002]. The excess shear stress, $\Delta \tau$, is defined as the difference between the shear strength and shear stress acting parallel to the fractures

$$\Delta \tau = \tau - \mu_d \sigma_n, \qquad (2)$$

where $\tau$ is the shear stress. For fracture faces that violate the Mohr–Coulomb criterion, the shear displacement, u, can be approximated by dividing the excess shear stress by the shear stiffness, $K'_s$, such as in linear elastic theory:

$$u = \frac{\Delta \tau}{K'_s}. \qquad (3)$$

We note that the static/dynamic friction model falls into the category of an 'inherently discrete model' because the strength of the failing elements drops discontinuously with the slip [*Rice*, 1993]. Nevertheless, the model has been reported to provide qualitatively acceptable results [*McClure and Horne*, 2011].

To describe the strength of the seismicity, the shear displacements and the fracture face areas that experience shear displacements are correlated to the seismic moment, $M_0$:

$$M_0 = \int G u dA, \qquad (4)$$

where G is the shear modulus and A is the fracture surface area associated with the shear displacement.

When the fracture faces start to slip, the characteristics of the opposing fracture surfaces are altered because of the asperity damage, which takes the form of irreversible increments in the void space [*Barton et al.*, 1985]. For flow simulation purposes, the void space between fracture faces is identified as the mechanical aperture and calculated by considering the fracture deformations and the mechanical aperture of the initial state. The final form of the mechanical aperture, E, can be written as follows:

$$E = E_0 - \Delta E_{rev} + \Delta E_{irrev}, \qquad (5)$$

where $E_0$ is the initial state and $\Delta E_{irrev}$ is the irreversible dilation of the fracture because of the shear slip, which is calculated by multiplying the tangent of the dilation angle, $\varphi_{dil}$, by the slip in the shear direction, u. The initial state is assumed to be the mechanical aperture measured under zero stress conditions and is further correlated with the porosity of the fracture because it can correspond to the conditions when the porosity of the fracture is equal to one. The change in the porosity of the fracture can be calculated accordingly.

Although the mechanical aperture is related to the porosity of the fracture, the permeability of the fracture is linked to the hydraulic aperture between fracture surfaces by a cubic law [*Jaeger et al.*, 2009]. The hydraulic aperture depends on several parameters, such as the roughness, tortuosity and contacts between the fracture surfaces [*Barton et al.*, 1985; *Olsson and Barton*, 2001]. These effects are lumped into the 'joint roughness coefficient' (JRC), which was introduced by Barton as follows:

$$e = \frac{E^2}{JRC^{2.5}} \qquad (6)$$

where e is the hydraulic aperture. JRC values can be obtained experimentally or via comparisons between the observed fracture surfaces and previously measured fracture surfaces [*Barton and Choubey*, 1977].

Although the elevated pressure results in an increase of the mechanical aperture in the injection phase, the opposite effect occurs during the post-injection phase according to equation (1) and equation (5) because of the reversibility of $\Delta E_{rev}$. As the mechanical aperture of a fracture is correlated with its permeability and porosity, the pressure drop around the injection region after the termination of the well reverses the increased porosity and permeability (to some extent) provided by the injection phase. Because less void space is observed compared with that in the injection phase, the fractures around the well act as a fluid source and push the pressure front further, thereby causing additional seismic activity.

2.2 Multi-physics coupling

Prior to the hydraulic stimulation, the reservoir is assumed to be stable under in situ conditions. The fractures are described as two surfaces that stay in contact (the effective normal loading is greater than zero) but still permit fluid flow because of microscale surface roughness. Our model includes fluid flow in the fractures and matrix, fracture deformation, and rock matrix deformation as well as the coupling of these three processes.

The modeled stimulation process is conducted by the injection of an isothermal, single-phase, slightly compressible fluid. The fluid flow is governed by the mass balance and Darcy's law both in the fractures and the matrix. Implicit time discretization is used for the time dependency of the governing equations. For the spatial discretization, the flow equations are approximated by a cell-centered finite volume method, specifically, the Two-Point Flux Approximation (TPFA). The fracture flow and fracture-matrix interaction are modeled by a Discrete Fracture Matrix (DFM) approach as implemented in the MATLAB Reservoir

Simulation Toolbox (MRST) [*Lie et al.*, 2012]. The DFM models [*Karimi-Fard et al.*, 2003; *Sandve et al.*, 2012] enable the modelling of reservoir fluid flow because dominant fractures are explicitly represented in an otherwise low-permeability rock. The region surrounding the explicitly represented fracture network accounts for low but non-negligible flow in the pores and small-scale fractures.

The shear displacements induced by the fluid injection result in stress alterations and deformations in the rock matrix. The perturbations in the stress state are governed by the equations of quasi-static equilibrium with negligible body forces, whereas the weight of the overburden is accounted for by the background stress conditions. The matrix is assumed to be a linear elastic body with fractures represented as slit-like discontinuities. The poroelastic effects in the matrix are neglected, and the fluid pressure on the fracture walls is coupled with the mechanical deformation. We use a recently developed cell-centered finite volume method, Multi-Point Stress Approximation (MPSA) [*Keilegavlen and Nordbotten*, 2017; *Nordbotten*, 2014], to approximate the solution of the linear momentum balance equation because it enables the use of the same data structure applied in the fluid flow discretization. MPSA enables the efficient inclusion of fractures by modeling them as internal boundary conditions, including fracture deformations in both the normal direction (change in the mechanical aperture) and the shear direction (caused by excess shear stress) [*Ucar et al.*, 2016].

To couple the fluid flow, fracture deformation and rock matrix deformation, a procedure that includes two iterations is used. For each time step, the first iteration is conducted between the pressure, the reversible mechanical deformation, which is $\Delta E_{rev}$ in equation (1), and the matrix deformation to capture the dynamics of the mechanical aperture change associated with the pressure and the response of the rock matrix. The iteration is terminated when a predefined threshold for the update in the mechanical aperture is met. After the convergence of this iteration, the second iteration is applied to capture the initiated shear displacements caused by the decreased friction resistance by checking the Mohr-Coulomb criterion for each fracture face. These shear displacements result in an irreversible contribution to the fracture deformation, which is $\Delta E_{irrev}$ in equation (5). Because of the stress alteration caused by the shear displacements, the Mohr-Coulomb criterion is rechecked for possible additional induced shear displacements, which are referred to as 'shear avalanches' [*Baisch et al.*, 2010]. The Mohr-Coulomb criterion check is conducted until stress alterations caused by slip no longer induce additional displacements, and then the solution continues with the next time step.

## 3 Simulation Study

Two three-dimensional case studies were designed to investigate the effect of the post-injection normal closure of fractures on the induced seismicity. The first case study presented a single fracture, and the second case study presented a complex fracture network.

### 3.1. Simulation setups

For the first simulation in case 1, a single fracture with a 1500-m diameter was examined. The fracture had an angle of 36° from the maximum horizontal principal stress direction and a 70° dip angle. The mechanical aperture under zero-stress conditions, $E_0$, was set as 2.25e-04 m, and the time step was 450 s. The injection started at a rate of 1.5e-05 kg/s and increased at each time step to 7.5e-05 kg/s, 4.2e-04 kg/s and 7e-04 kg/s. The seismic moment release after a half-hour injection was examined. For the second simulation in case 2, a connected fracture network with 20 fractures of diameters ranging between 500 m and 1500 m was assessed. The geometric details of the fracture network can be found in Table 1. The mechanical aperture under zero-stress conditions, $E_0$, was set as 2e-04 m except for the fracture that included the source term, where a higher $E_0$ of 9e-04 m was used to allow for higher injection amounts. The same time step used in case 1 (i.e., 450 s) was applied. The injection started at a rate of 1 kg/s and increased at successive time steps of 2 kg/s, 3 kg/s and 3 kg/s. For simplicity, the injection well was modeled as a source term inside a fracture. Because the injection is conducted from the small volume of a single source cell to the fractures and to the low-permeability matrix, low injection rates were used to avoid exceeding the minimum principal stress. For these conditions, the fluid flow mainly occurred in the fractures, while leakage into the matrix was not neglected for either case.

| Fracture No | x (m) | y (m) | z (m) | Fracture Diameter (m) | Strike | Dip |
|---|---|---|---|---|---|---|
| 1 | 0 | 0 | -5000 | 1500 | 100 | 80 |
| 2 | -200 | -200 | -4850 | 700 | 170 | 70 |
| 3 | 500 | 200 | -5200 | 850 | 130 | 70 |
| 4 | 100 | 100 | -5000 | 1000 | 60 | 80 |
| 5 | -400 | 100 | -5500 | 800 | 100 | 30 |
| 6 | -400 | 300 | -5000 | 700 | 170 | 70 |
| 7 | -600 | -100 | -4600 | 700 | 50 | 70 |
| 8 | -500 | 0 | -5000 | 700 | 150 | 85 |
| 9 | -500 | 0 | -5700 | 1200 | 300 | 45 |
| 10 | -400 | -200 | -5900 | 800 | 300 | 80 |
| 11 | -400 | -100 | -5700 | 800 | 100 | 50 |
| 12 | 400 | 400 | -5500 | 1000 | 250 | 85 |
| 13 | -600 | -100 | -4200 | 700 | 220 | 75 |
| 14 | 0 | 0 | -4400 | 700 | 300 | 25 |
| 15 | 500 | 200 | -5800 | 700 | 300 | 25 |
| 16 | 700 | 100 | -5800 | 700 | 100 | 75 |
| 17 | 0 | 300 | -4300 | 500 | 150 | 75 |
| 18 | 0 | 300 | -5000 | 500 | 120 | 75 |
| 19 | -300 | -600 | -4900 | 700 | 300 | 25 |
| 20 | -200 | -900 | -4900 | 600 | 30 | 75 |

**Table 1.** Details of the fracture network used for case 2. The coordinates of the center of the fractures (x, y, and z) are calculated by considering the wellhead of fracture no 1 (main fracture) as the origin.

The remaining parameters were common to both simulations and are listed in Table 2. The selected computational domains were sufficiently large so that the boundary effects were negligible. The computational grids were constructed with an unstructured tetrahedron mesh that has approximately 10000 triangles, which was used to discretize the fractures, and 40000 cells, which was used to discretize the domain for both cases; thus, the single-fracture case has a finer grid. We used different mesh sizes in our simulations to show the normal closure effect for different grid sizes and avoid the high computational costs associated with the larger fracture network.

| Fluid Flow Parameters | |
|---|---|
| Initial hydrostatic pressure | 40 MPa |
| Fluid initial density | 1014 kg/m$^3$ |
| Fluid compressibility | 0.000458 MPa$^{-1}$ |
| Rock matrix permeability | 1nD |
| Rock matrix porosity | 0.01 |
| Fracture Parameters | |
| Normal stiffness, $K_n$ | 3.5x10$^{16}$ N/m |
| Shear stiffness per area, $K'_s$ | 2 GPa/m |
| Dilation angle | 3 |
| JRC | 15 |
| Maximum possible closure, $\Delta E_{max}$ | $E_0$ |
| Friction, $\mu_s$ and $\mu_d$ | 0.6 and 0.55 |
| Elastic Parameters of Rock Matrix | |
| Shear modulus, G | 21.6 GPa |
| Poisson's ratio | 0.25 |
| Principle stresses ($S_H$, $S_h$, $S_V$) | 120 MPa, 80 MPa, 100 MPa |

**Table 2.** The common parameters used for both case 1 and case 2.

For both of the fracture networks, the effect of the normal closure of the fractures was analyzed by conducting stimulation scenarios with and without the inclusion of the normal closure of the fractures after the termination of the injection. While one simulation scenario applied the modeling approach described above, the other kept $\Delta E_{rev}$ constant in the post-injection period so that only the normal deformation occurred through shear dilation, $\Delta E_{irrev}$.

Notably, if the normal closure effect was eliminated from the beginning of the injection, almost all characteristics (pressure distribution, fracture permeability/porosity, stress conditions etc.) of the compared systems would differ significantly when the injection stopped. Therefore, we kept the scenarios identical for the injection phase to ensure a fair comparison of the results, and the normal closure effect was only isolated in the post-injection phase.

3.2 Case 1: single fracture

In this case, the mechanical aperture change and the post-injection seismicity of the single fracture were examined with and without the normal closure effect. Figure 1 depicts the spatial distribution of the mechanical aperture change and the seismicity of the fracture during and after the injection phase. At the end of the injection, the mechanical aperture change was highest at the injection cell and slowly decreased towards the boundary of the stimulated region. Figure 1b shows the change in the mechanical aperture after the shut-in of the well when the aperture is allowed to change because of the normal closure. The darker region in the middle shows negative change values for the mechanical aperture, which indicated that the fracture closure was caused by the pressure decrease around the well. Figure 1c illustrates the mechanical aperture change in the simulation scenario when the fracture was not allowed to close after the shut-in of the well. The middle region in this figure shows no change in the mechanical aperture. A comparison of Figures 1b and 1c clearly shows the effect of the normal closure. The width of the region of the increased mechanical aperture was wider for the case that included normal closure (Figure 1b) because of the extra fluid flow caused by the decreased porosity in the center region. Moreover, Figure 1d shows the released seismic moments for the two simulation scenarios according to the induced shear displacements. Both scenarios created the same seismicity for the injection phase as modeled; however, the normal closure effect induced a higher seismic moment in the post-injection phase. This moment was the direct result of the enhanced propagation of the pressure front by the normal closure of the fracture, which again decreased the frictional resistance over a wider area, thereby creating additional seismicity. The total seismic moments with and without the normal closure effect were calculated as 2.99e+12 Nm and 2.28e+12 Nm, respectively.

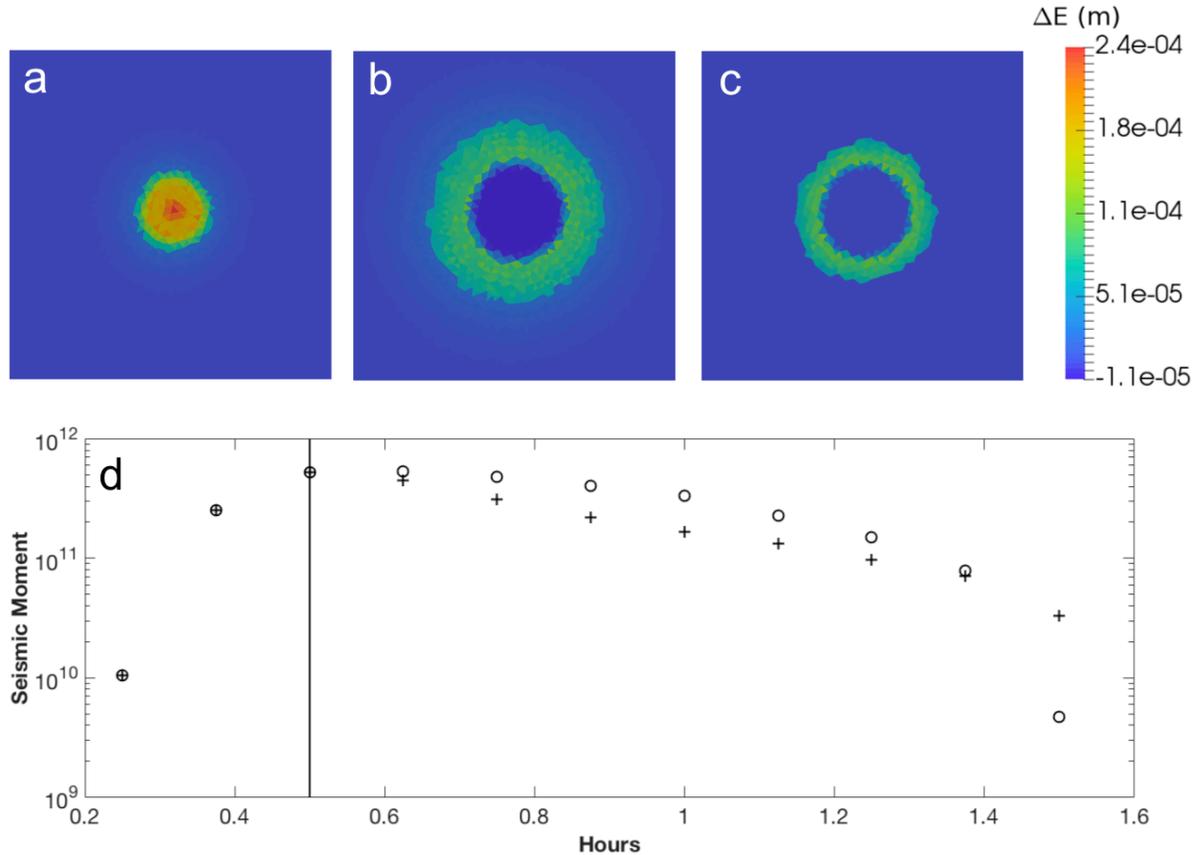

**Figure 1**. Normal closure behavior of the single-fracture case. Mechanical aperture changes that accumulated during (a) injection, (b) post-injection with the normal closure effect, and (c) post-injection without the normal closure effect. (d) Moment of the induced seismicity as a function of time. Circles and crosses show the modeled seismicity with and without the normal loading effect, respectively. The black solid line indicates the time for the termination of the injection.

3.3 Case 2: complicated fracture network

The normal closure effect in a complicated three-dimensional fracture network was also examined with and without eliminating the fracture closure after the termination of the injection, such as in case 1. The main results of these numerical examples, which are the mechanical aperture change and the induced seismic moments during the injection and in the post-phase are presented in Figure 2. For Figure 2, the pressure distribution and mechanical aperture change can vary throughout the fractures. Thus, the pressure is not constant throughout the entirety of the fractures; however, these differences are not easily observed in the figure because of limitations caused by the large range of the color bar. The highest aperture changes are observed at the fracture where the injection is occurring, which is located in the approximate middle of the domain. We refer to this fracture as the "main fracture" (marked with a circle in Figure 2 for clarity). The total mechanical aperture change of the main fracture for the case without normal closure after injection was higher than that for the case with normal closure, which was expected. For the case with normal closure, the change in the mechanical apertures achieved a peak value immediately before the injection stopped and then decreased after the termination of the injection. The shut-in of the injection source caused a pressure decrease primarily around the injection region (in other words, in the main fracture), which reduced the porosity and caused an additional flux and advancement of the pressure front. These changes created significant induced seismicity for fractures far from the injection source, which is consistent with the observed seismicity in the Basel and Paralana geothermal fields. Because a complex fracture network that includes interactions between several fractures with completely different orientations was used, the pressure front reaches favorably oriented fractures far from the injection region. For the scenario in which the normal closure was excluded, less induced seismicity was observed because the pressure front could not reach favorably oriented fractures. The total generated seismic moments with and without the normal closure effect were calculated as 2.47e+14 and 9.09e+13 Nm, respectively.

The lower part of Figure 2 shows the released seismic moment before and after the injection phase for each fracture. Although the seismicity was triggered for only four fractures when the normal closure effect was disregarded, the number of fractures exposed to the induced seismicity reached 7 when the normal loading effect was included. For the injection phase of the cases, induced seismicity was observed for four fractures (shown by the markers o, +, x, and □ in Figure 2). For the post-injection phase, the same four fractures continued to generate seismicity for the case without normal loading. However, when the normal closure effect was included in the model, three different fractures that were not activated during injection started inducing seismicity in the post injection phase. For example, the fracture marked by ◊ in Figure 2 started to induce seismicity almost two hours after the termination of injection and continued to generate seismicity for five hours. Moreover, the total time of the induced seismicity was two hours longer for the scenario in which the normal closure was included than the total time of the scenario in which the normal closure was not included. Note that for both scenarios, certain fractures were not exposed to any aperture change because of their orientation. If the fracture orientation was not favorable in terms of the aperture (see equation (1)) or stress (Mohr-Coulomb criterion), the pressure change for those fractures was limited, which led to a lack of observations of induced seismicity.

The current modeling approach enables us to analyze several fracture networks that have different characteristics, such as the number, shape and orientation of fractures and the network connectivity. Therefore, after validating the effect of the normal closure of fractures on the induced seismicity, we performed further studies to investigate seismic events that occur in the post-injection phase. The extent and magnitude of the post-injection-induced seismicity were significantly dependent on the number of favorably orientated fractures and the connections among them. The friction coefficient for all fractures was assumed to be identical in the results presented here, although it will differ for each fracture in a real fracture network. When we used different friction coefficients for different fractures, large seismic moments were generated once the pressure front reached a particularly favorably oriented fracture with low friction. Moreover, the current friction modeling limited the accuracy of the seismicity characteristics because of the discretization dependency [*Rice*, 1993]. Furthermore, several assumptions, such as a constant stiffness coefficient, and the removal of poroelastic effects from the matrix as well as thermoelastic and geochemical effects limited the accuracy of the results. However, these assumptions would not interfere with the effect of the normal closure of fractures on the post-injection seismicity.

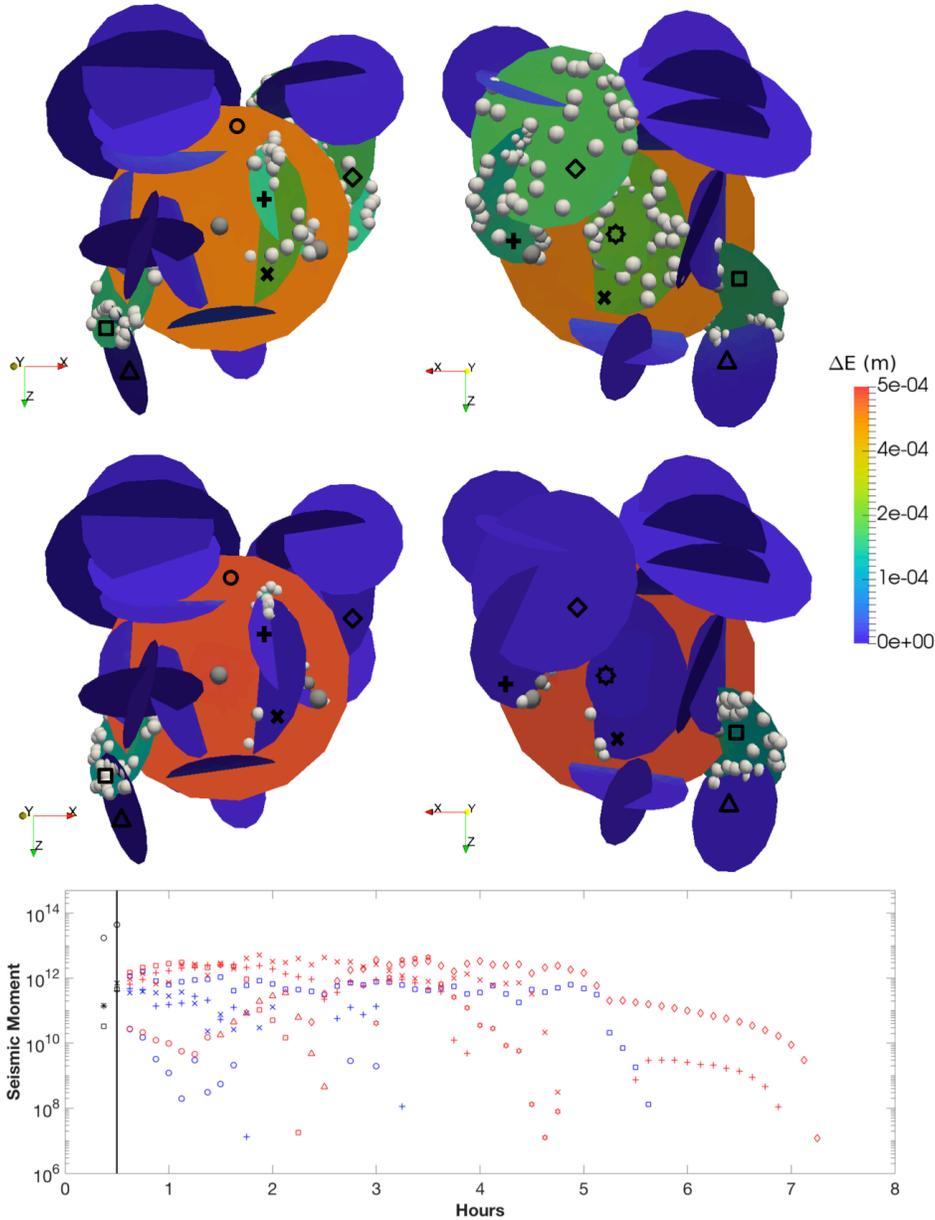

**Figure 2**. Top and middle figures: Seismicity and mechanical aperture change (ΔE) in a complicated fracture network. Modeled seismic events with the normal loading effect (top) and without the normal loading effect (middle) shown by two different views. The spheres, which are associated with the seismic moments, are located on the faces where the largest event occurred for each fracture. The radiuses of the spheres are correlated with the magnitude of the seismic moment. The gray spheres show the seismic moments that occurred during the injection phase, and the white spheres indicate the moments that occurred after the termination of the injection. Bottom figure: Moment of the induced seismicity as a function of time. Each marker type represents a fracture as indicated in the top and the middle figures. Red markers show the modeled seismicity for the scenario in which the normal loading effect is included in the model after shut-in, and blue markers show the results for the scenario in which fracture closure is disregarded in the model after shut-in. The black solid line indicates the time for the termination of the injection.

# 4 Conclusions

The normal closure of fractures as a mechanism for induced seismicity in the post-injection phase of hydraulic stimulation is examined. Two different 3-D situations are studied: a reservoir with a single fracture and a complicated fracture network that includes 20 fractures. A detailed analysis of the underlying mechanism is performed by comparing the results from numerical experiments conducted with and without the elimination of the normal closure effect. A qualitative examination of the results for both types of fracture network suggests that the normal closure of fractures should be considered a significant mechanism that leads to a large post-injection seismicity. Furthermore, the results indicated that the structure of the complex fracture network strongly affects the evolution of the induced seismicity, and the location of the induced seismicity modeled in this network showed analogous characteristics with recorded seismicity from stimulation of geothermal reservoirs.

## Acknowledgments

Data associated with this study can be obtained upon request by contacting the corresponding author .The work was funded by the Research Council of Norway through grant no. 228832/E20 and Statoil ASA through the Akademia agreement.